\documentclass[twocolumn,showpacs,preprintnumbers,amsmath,amssymb]{revtex4}
\usepackage{graphicx}% Include figure files
\usepackage{dcolumn}% Align table columns on decimal point
\usepackage{bm}% bold math

\usepackage{tabulary}
\usepackage{psfrag}

\begin{document}

%\preprint{APS/123-QED}

\title{Infrared Gluon and Ghost Propagator Exponents From Lattice QCD}
% Force line breaks with \\

\author{O. Oliveira, P. J. Silva}
\affiliation{Dep. F\'{\i}sica, Universidade de Coimbra, 3004-516 Coimbra, 
             Portugal}%

\date{\today}% It is always \today, today,
             %  but any date may be explicitly specified

\begin{abstract}
The compatibility of the pure power law infrared solution of QCD 
and lattice data for the gluon and ghost propagators in Landau gauge is discussed. 
For the gluon propagator, the lattice data is well described by a pure power law with an infrared exponent $\kappa \sim 0.53$, in the Dyson-Schwinger notation. $\kappa$ is
measured using a technique that suppresses finite volume effects.
This value implies a vanishing zero momentum gluon propagator, in agreement with the
Gribov-Zwanziger confinement scenario.
For the ghost propagator, the lattice data seem not to follow a pure power law, at least for
the range of momenta accessed in our simulation.
\end{abstract}

                             % PACS, the Physics and Astronomy
                             % Classification Scheme.
%\keywords{Gluon Propagator, Confinement, Infrared QCD}
                %Use showkeys class option if keyword
                %display desired
\maketitle

%=======================================================================
\section{\label{intro}Introduction and Motivation}

The infrared properties of the Landau gauge gluon and ghost propagators in 
momentum space, respectively,
\begin{eqnarray}
  D^{ab}_{\mu\nu} (q) & = & \delta^{ab} \, 
           \left( \delta_{\mu\nu} \, - \, \frac{q_\mu q_\nu}{q^2} \right)
           \, D(q^2) \, ,  \\
  G^{ab} (q) & = & - \delta^{ab} G(q^2) \, ,
\end{eqnarray}
are connected with gluon confinement mechanisms, namely the Kugo-Ojima scenario (KO) \cite{KuOj} and the Gribov-Zwanziger horizon condition (GZ) \cite{Gribov,Zw91}.  
The GZ mechanism requires $D(0) = 0$ (which implies maximal violation of reflection positivity) and an enhanced ghost propagator, relative to the perturbative function. The KO confinement mechanism demands $1 / (q^2 G(q^2)) =  0$ in the limit $q \rightarrow 0$. From the point of view of the KO and GZ confinement mechanisms, the requirements on $D(0)$ and $G(0)$ are necessary conditions and its violation immediatly rules out these scenarios.

In the recent years there has been a renewed interest in the computation of gluon and ghost propagators in the pure gauge theory, due to progress on solutions of the Dyson-Schwinger equations (DSE) and lattice simulations which explore further the infrared region. 

The DSE are an infinite tower of coupled nonlinear equations for the QCD Green's functions. The computation of the gluon and ghost propagators within DSE requires defining a truncation scheme and parametrizing some of the QCD vertices. For the Landau gauge, recent solutions can be classified in two
 categories: solutions which are not compatible with the KO or GZ confining mechanism \cite{AgNa03,AgNa04,Ag07,Ag08}
and solutions which do not rule out the two confining mechanisms \cite{Fischer06,Fischer03,Fischer02,Fischer02a}. For the first class of solutions,
 $D(0)$ is finite and does not vanish.  For the later class of solutions, in \cite{LeSm02} an analytical solution was found for the deep infrared region \cite{Review}. The solution assumes infrared ghost dominance and connects the two propagators via a single exponent, $\kappa$,
\begin{eqnarray}
  Z( q^2 ) ~ = ~ q^2 \, D( q^2) & = &
         \omega \left( \frac{q^2}{\sigma^2} \right)^{2 \kappa} ,
  \label{Zdse} \\
  F( q^2 ) ~ = ~ q^2 \, G( q^2) & = &
           \omega^\prime \left( \frac{q^2}{\sigma^2} \right)^{- \kappa} ;
  \label{Fdse}
\end{eqnarray}
$\sigma$ is a constant with dimension of mass. Moreover, DSE equations predict $\kappa = 0.595$, which, for the zero momentum, implies a null (infinite) gluon (ghost) propagator. Furthermore, renormalization group analysis \cite{Pa04,FiGi04} restrict the possible values for $\kappa$ to 
$0.52 \, \le \, \kappa \, \le \, 0.595$. This result suggests a null (infinite) zero momentum gluon (ghost) propagator. 
In \cite{FisPaw07} it was argued that the solution (\ref{Zdse})-(\ref{Fdse}) is the unique power law infrared solution compatible 
with DSE and functional renormalization group equations. In \cite{Zw02}  a similar analysis of the DSE within time-independent stochastic quantisation predicted the same behaviour and $\kappa = 0.52145$.  

The computer simulations of 4D pure SU(3) and SU(2) gauge theories on a lattice
\cite{Leinweber,Furui,Berlim,Braz,su2vsu307,Stern07,Bo08,qnp,
Cucc07,Ol07,Bo07,Cu07,
OlSi08,Maas08,Oliveira05,Bonnet01,Bloch04,Boucaud06}
%PAULO
% acrescentei OlSi08,Maas08,Oliveira05,Bonnet01,Bloch04,Boucaud06}
%PAULO
 show a finite nonvanishing zero momentum gluon propagator.
Although $D(0)$ is a decreasing function of the lattice volume, from naive extrapolations to the
infinite volume \cite{Cu07,Oliveira05,Bonnet01} it is not clear if the lattice result becomes compatible
with the GZ confinement mechanism, i.e. if $D(0)$ approaches zero as $V \rightarrow \infty$. 
Moreover, recent 4D lattice QCD calculations of the gluon propagator on large volumes,
$(27$ fm$)^4$ for SU(2)  \cite{Cucc07} and $(13$ fm$)^4$ for SU(3) \cite{Bo07}, show a propagator 
which goes to a
constant in the infrared limit, although not excluding a gluon propagator tending towards zero. However,
to work out such huge lattice volumes, the simulations were carried out with the Wilson action and
using a relatively large value for the lattice spacing, $a \sim 0.21$ fm for SU(2) and $a \sim 0.165$ for
SU(3). It is not clear yet whether this have any influence on such
propagator.

Recently, in \cite{Cu07} rigorous upper and lower bounds for the zero-momentum gluon propagator
were derived and its scaling behaviour with the volume analyzed for the SU(2) Yang-Mills theory.
The authors found a finite and non-vanishing $D(0)$ for SU(2) in the infinite volume limit. 
In \cite{OlSi08}, the same analysis was performed but for the SU(3) Yang-Mills theory. It turns out that
for SU(3), the infinite volume limit gives a $D(0) = 0$. The different behaviours in the infinite volume 
limit is a puzzling result and, presently, we have no understanding for such a difference.

In what concerns the lattice ghost propagator 
\cite{Berlim,Braz,su2vsu307,qnp,Bloch04,Boucaud06,Maas08}, the simulations show an enhanced propagator but not in agreement with that predicted by the solution (\ref{Fdse}).  The lattice data seems
to be closer to the solution of \cite{Ag08}. Indeed, the lattice data shows a propagator which is almost 
identical to the perturbative propagator. 

In order to try to understand the difference between the DSE solution which compatible with the
GZ and KO confinement mechanisms, in \cite{DSEtorus} the DSE were solved on a 4D symmetric torus. 
According to the authors, the gluon and ghost propagators approach slowly the infinite volume value. Moreover, they claim that to observe the suppression of the gluon propagator one should go to volumes
as large as $(10$ fm$)^4$. Note that recent lattice simulations  \cite{Cucc07,Bo07} have volumes well above  the $(10$ fm$)^4$.

In \cite{Braz,
%PAULO
qnp, Oliveira05,NosSard,PRD06} we have tried to measure both the gluon and ghost propagators using a set of large asymmetric lattices, i.e.  $L^3 \times T$ with $T$ larger than $L$, to access the deep infrared
region and to check the compatibility between the lattice data and the solution (\ref{Zdse})-(\ref{Fdse}). 
The lattices used are larger than 10 fm, by a factor of $\sim 2.5$, in the temporal direction and are much
shorter, by a factor of $\sim 1/5$, in the spatial directions. 
For the $L^3 \times T$ lattices, besides the finite volume effects also observed in simulations with symmetric $L^4$ lattices, one has to care about how the asymmetry changes the gluon and ghost propagators data. The finite lattice effects in $D(q^2)$ and $G(q^2)$ in the asymmetric lattices are, qualitatively, equal to the effects observed in the solutions of the DSE on a symmetric 4D torus \cite{DSEtorus} and on the simulation of 3D asymmetric SU(2) lattices \cite{CuMe06}. 

For the gluon propagator, in \cite{PRD06}, using the lattice dressing function $q^2 D(q^2)$, and excluding the zero momentum point from the analysis,  we have demonstrated that the gluon lattice data
is compatible with (\ref{Zdse}). 
We would like to call the reader attention that we started by simulating $16^3 \times 128$ and 
$16^3 \times 256$ lattices for $\beta = 6.0$. For these two lattices, it turned out that the 
$16^3 \times 128$ data cannot be described by a pure power law. 
Only the fits to the $16^3 \times 256$ gluon data have acceptable $\chi^2/d.o.f.$, i.e.
$\chi^2/d.o.f. <  2$. 
Furthermore, an attempt to extrapolate the lattice to infinite spatial volume suggests a $\kappa$ in the
range 0.498 to 0.525. Despite this result, which gives some support to the KO and GZ confinement
scenarios, before extrapolations the lattice data shows no suppression of the gluon propagator for 
small momenta, except for $D(0)$ when compared to the first non-zero momentum (see figure
\ref{GlueProp}).  

In what concerns the ghost propagator computed using asymmetric lattices \cite{Braz,qnp}, 
we have observed that $G(q^2)$ is enhanced, compared to the perturbative solution, in the infrared
region (see figure \ref{GlueProp}), but not as much as predicted by the DSE solution (\ref{Zdse}).

In this work we discuss a method which, assuming a pure power law behavior for the infrared propagators, aims to measure the infrared exponents without relying on data extrapolation. The
measure relies on the definition of a convenient ratio of propagators which, in principle, suppresses the
finite volume effects. The results reported show that, for the same asymmetric lattices used previously,
the method provides estimates of the gluon propagator exponent which are stable against variation of
the range of momenta and variation of the spatial lattice extent $L$. Furthermore, the measured
$\kappa$ are compatible with the GZ confinement scenario, i.e. they predict a vanishng zero momentum
gluon propagator. On the other hand, the data for the ghost propagator, although  stable against variation of the lattice volume, is not compatible with a pure power law behaviour. 

The paper is organized as follows. In section \ref{details} we report on the asymmetric lattices used
to investigate the infrared gluon and ghost propagators. In section \ref{ratiosglsec} the method used to
measure the infrared exponents is discussed together with the results for the gluon propagator. The
section also includes a discussion of the finite volume and asymmetry effects. In section
\ref{ghost}, the method is applied to the ghost propagator and finally in section \ref{conclusions} we 
draw the conclusions.

%======================================================================
\section{Details of the Simulations \label{details}}

In this article we use Wilson action, $\beta = 6.0$, gauge configurations for the lattices reported in 
table \ref{Uvol}. The asymmetric lattices have a temporal extension of $T=256$ ($\sim 26$ fm) 
which allow us to access momenta as low as 48 MeV. 
The main difference to \cite{PRD06} being the larger statistics for the largest lattices. 

%=====================================================
%=====================================================
\begin{table}
\caption{Lattice setup. All simulations use a Monte Carlo sweep of 7 overrelaxation 
updates with 4 heat bath updates. The number of thermalization (Therm)
and separation (Sep) sweeps refers to the
combined sweeps. See \cite{PRD06} for details.} \label{Uvol}
\begin{ruledtabular}
\begin{tabular}{rrrr}
   Lattice           & Therm. 
                     & Sep.
                     & \# Conf. \\
\hline
   $8^3 \times 256$   & 1500 & 1000 & 80  \\
   $10^3 \times 256$  & 1500 & 1000 & 80  \\
   $12^3 \times 256$  & 1500 & 1000 & 80  \\
   $14^3 \times 256$  & 3000 & 1000 & 128 \\
   $16^3 \times 256$  & 3000 & 1500 & 155 \\
   $18^3 \times 256$  & 2000 & 1000 & 150 \\
\hline
\end{tabular}
\end{ruledtabular}
\end{table}
%==================================================
%=================================================

The propagators were computed in the minimal Landau gauge and the gauge fixing was performed
using a Fourier accelerated steepest descent algorithm; see \cite{PRD06} for details and definitions.

In the following, we will discuss the infrared propagators and will consider only time-like momenta, 
defined as
\begin{equation}
  q[n] \, = \, q_4 [n] \, = \, \frac{2}{a} ~ \sin \Big( \frac{\pi n}{T} \Big)
 \, ,
  \hspace{0.5cm}  n ~ = ~ 0, \, 1, \, \dots \, \frac{T}{2} \, ,
\end{equation}
where $T$ is the time lattice extent. For the conversion to physical units we use $a^{-1} = 1.943(47)$ GeV \cite{LatSpa}. 

\begin{figure}                                   
\includegraphics[scale=0.33]{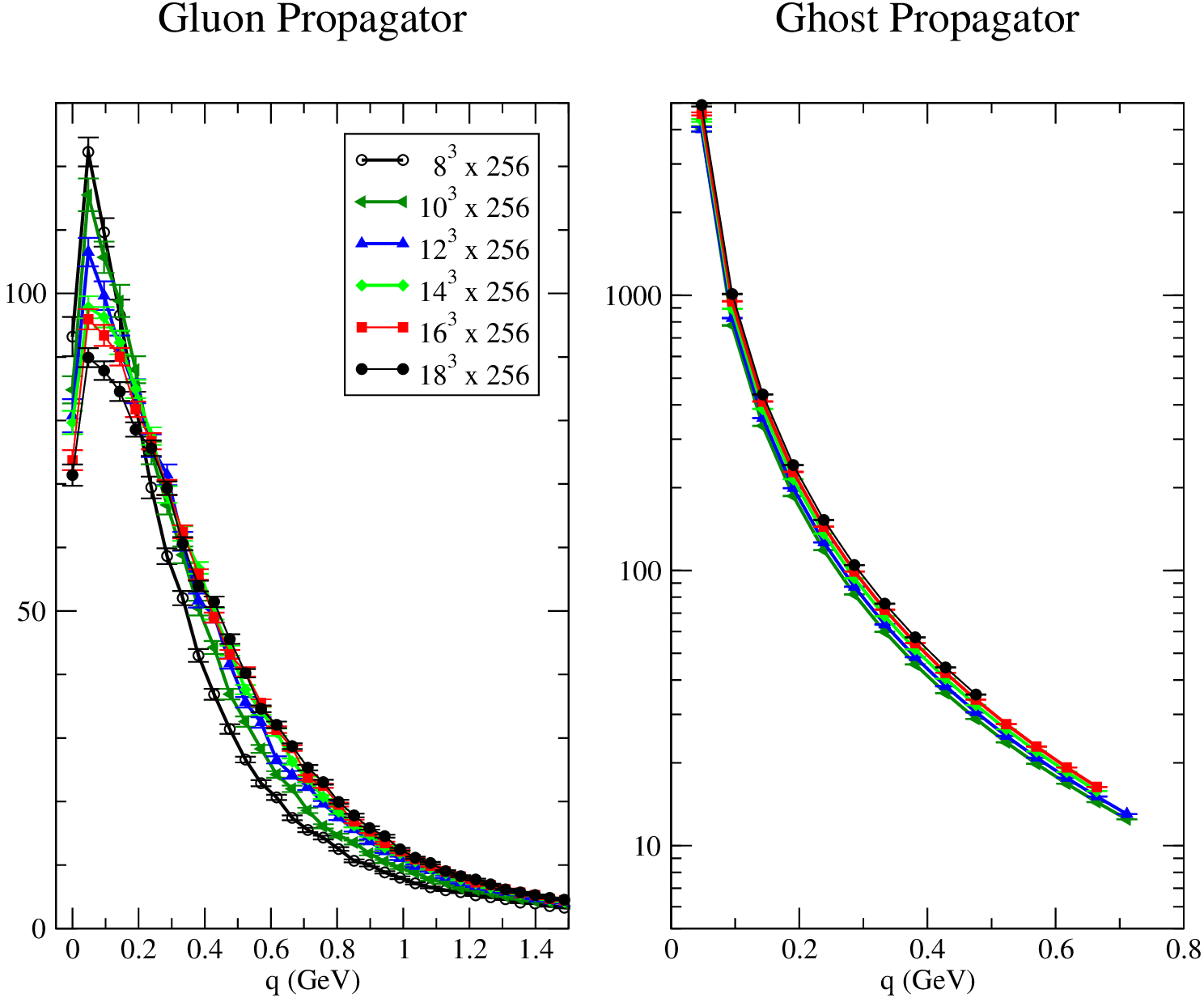}
\caption{\label{GlueProp} Bare gluon and ghost propagators 
for time like momenta. Note the logarithmic scale for the ghost propagator.}
\end{figure}

The ghost propagator was computed with the method described in \cite{Cucchieri97}, for the smallest
$q[n]$. In the calculation of the $D(q^2)$ and $G(q^2)$, the statistical errors were evaluated with the
jackknife method. Otherwise, the statistical errors were computed using the bootstrap method with 
a 68\% confidence level. The bare lattice gluon and ghost propagators are reported in 
figure \ref{GlueProp}. 

%========================================================
%=================================================

%======================================================================
\section{The Gluon Propagator\label{ratiosglsec}}

For the measurement of the infrared exponents, it will be assumed that the 
%PAULO
lattice 
%%%%%
dressing functions
$Z(q^2)$ and $F(q^2)$ are described by pure power laws $( q^2 )^\alpha$, 
as in (\ref{Zdse}) and (\ref{Fdse}), times a factor 
%PAULO
$\Delta ( q)$
% não será melhor para evitar confusões de notação com o apêndice?
%antes estava: $\Delta ( q^2, L, T)$ 
which summarises the finite volume corrections and/or deviations from the pure power law. If these corrections are constant (small), they are eliminated (suppressed) by taking ratios of $Z$
and $F$ at consecutive lattice momenta, i.e.
\begin{equation}
  \ln \left[ \frac{ \mathcal{G} (q^2[n+1])}{ \mathcal{G}(q^2[n])} \right] ~  = ~  \alpha
            \ln \left[ \frac{q^2[n+1]}{q^2[n]} \right]  ~ + ~ \cdots \, ,
            \label{ratiosdressing}
\end{equation}
where for the gluon $\mathcal{G}  (q^2) = Z (q^2)$ and, according to (\ref{Zdse}), $\alpha =  2 \kappa$.
For the ghost $\mathcal{G} (q^2) = F (q^2)$ and $\alpha = - \kappa$. In the infrared one expects that
corrections to (\ref{ratiosdressing}) are subleading.
Defining 
\begin{eqnarray}
R_{Z}[n]  & \equiv  & \ln \left[ \frac{Z(q^2[n+1])}{Z(q^2[n])} \right] \, ,  \nonumber \\
R_{q}[n]                 & \equiv   & \ln \left[\frac{q^2[n+1]}{q^2[n]} \right],
\label{def_rat}
\end{eqnarray}
we get, for the gluon propagator,
\begin{equation}
 R_{Z}[n]   =   2  \kappa ~ R_{q}[n].
\label{glueratio}
\end{equation}

%========================================================
\begin{figure}                                   
\includegraphics[scale=0.35]{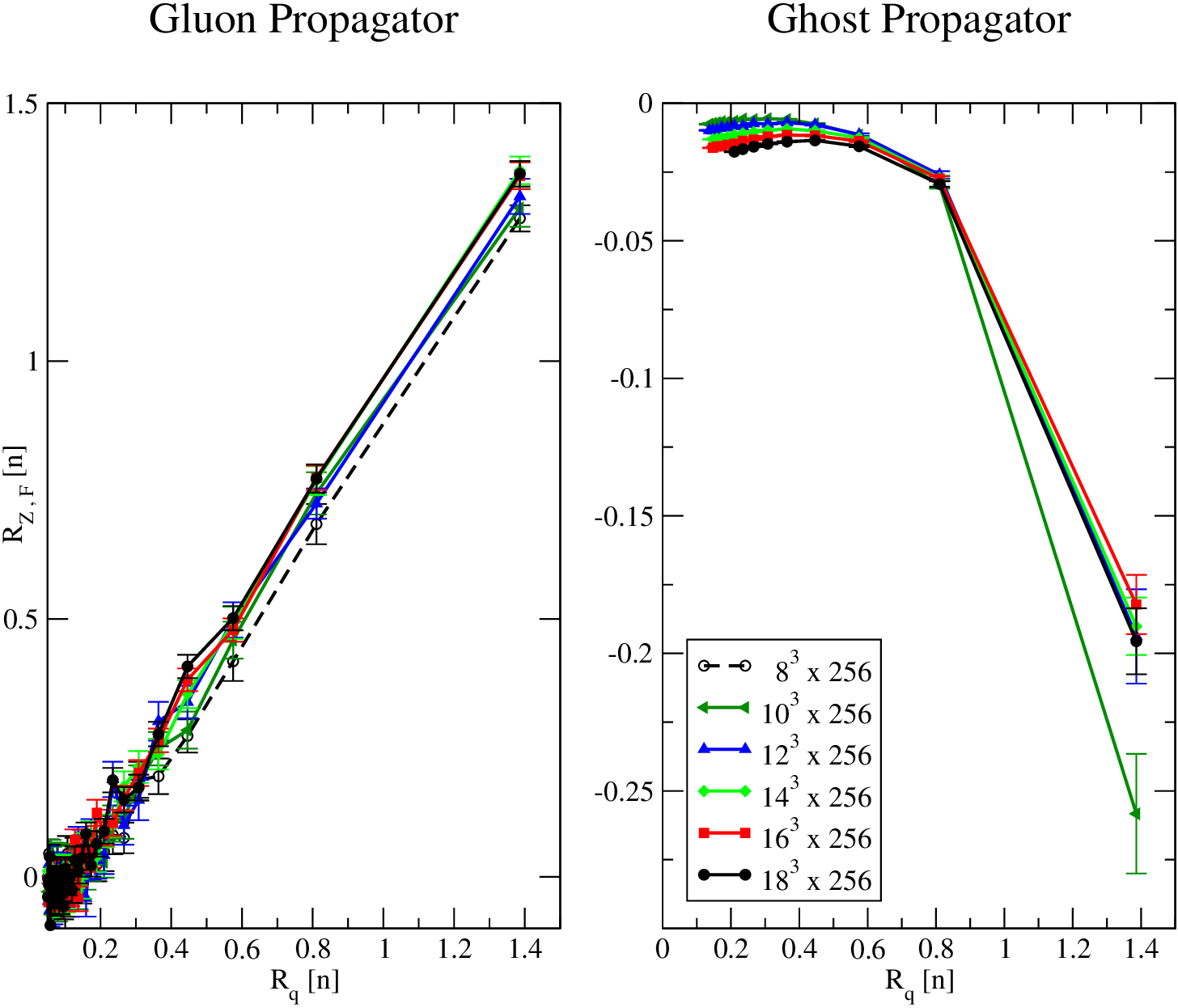}
\caption{\label{plotratios} Ratios of gluon (left) and ghost (right) dressing
functions. The statistical errors were computed using 1000 bootstrap samples 
for $8^3 - 12^3 \times 256$ lattices and 
1800 bootstrap samples for $14^3 - 18^3 \times 256$ lattices.}
\end{figure}
%================================================

The gluon data for $R_{Z}[n]$ as a function of $R_{q}[n]$, see figure \ref{plotratios}, shows a 
linear behavior for a surprising large range of momenta for all the lattices $8^3 - 18^3 \times 256$. 
Moreover, the slopes seem to be similar for all lattices. It seems that the corrections to (\ref{Zdse})
not suppressed by the ratios show up as a constant which adds up to (\ref{glueratio}). This hypothesis 
can be tested fitting the ratios to
\begin{equation}  
R_{Z}[n] ~  = ~  2  \kappa \, R_{q}[n] ~ + ~ C ~,
  \label{gluefit}
\end{equation}
assuming that $C$ is a constant. The fits of the lattice ratios to (\ref{gluefit}) using momenta up to 
$\sim 400$ MeV are reported in table \ref{tablegluefit}. 

The measured $\kappa$'s are stable against variation of the fitting range and spatial lattice size. 
Note that the $\kappa$ measured with the smallest lattice $8^3\times 256$ is compatible with the 
$\kappa$ measured using our largest volume $18^3\times 256$. Moreover, the central values for 
$\kappa$ are clearly above 0.5 and 
typically, within one standard deviation, $\kappa \ge 0.5$. In particular,
for the three largest volumes $14^3 - 18^3 \times 256$, which have the largest statistics, 
$\kappa > 0.5$ within one standard deviation. Statistical errors on $\kappa$ decrease as the fitting 
range increases, because one is using a larger set of data. For the largest fitting range ($q<381$ MeV)
and for the three largest lattices, the $\kappa$ are compatible with 0.5 only within 4 standard deviatons.
In this sense, in what concerns the infrared gluon propagator, the fits to (\ref{gluefit}) point towards 
$\kappa \sim 0.53$, suggesting a vanishing gluon propagator at zero momentum.

\begin{figure}[t]
\vspace*{0.2cm}                                  
\includegraphics[scale=0.35]{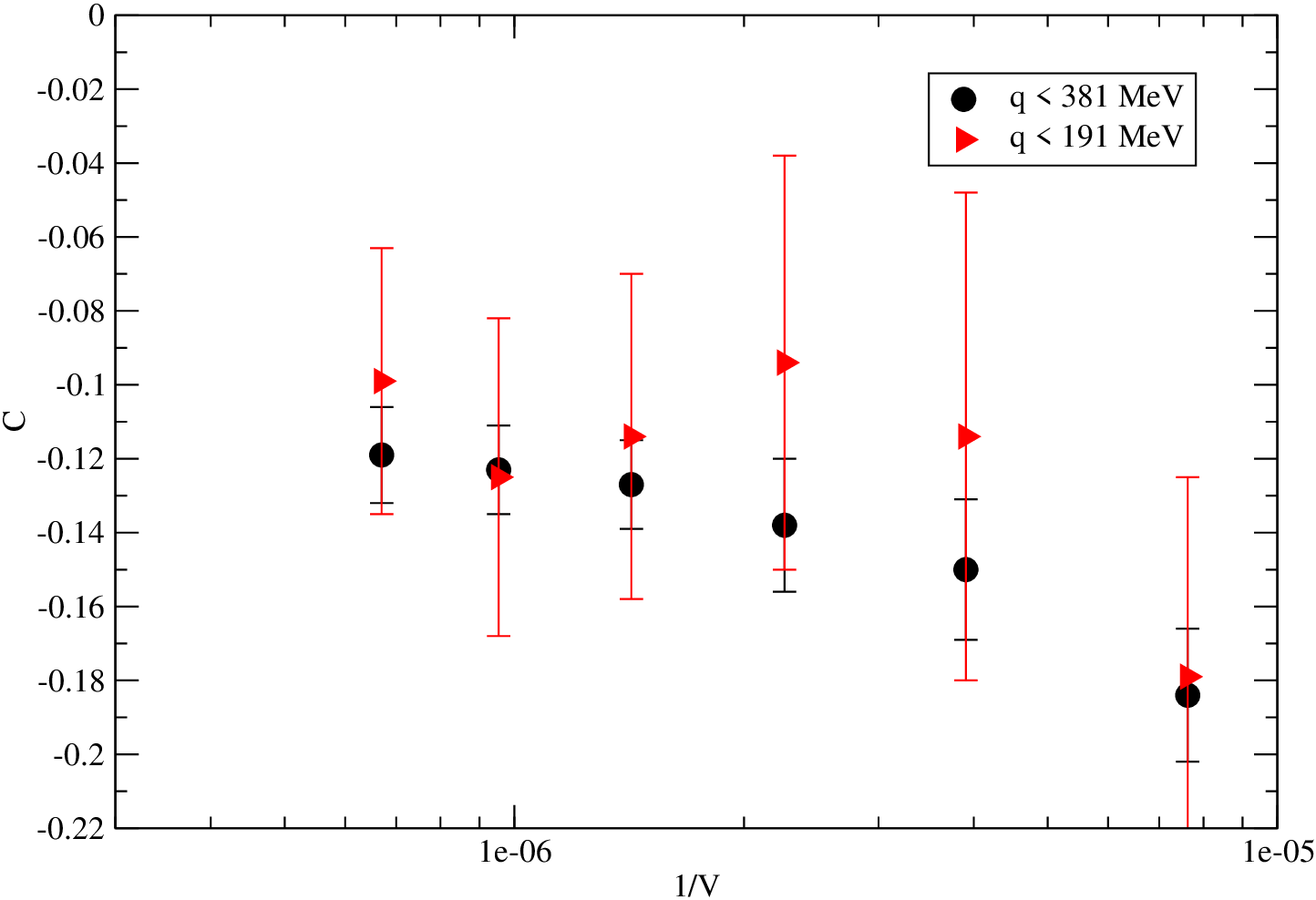}
\caption{\label{cplot} $C$ as function of $1/V$ for asymmetric lattices with $T=256$.Values were taken from table \ref{tablegluefit} considering the largest fitting range $q<381$ MeV and the smallest $q<191$ MeV.}
\end{figure}

Note that the absolute value of the constant C, in general, approaches zero as the lattice volume increases (see figure \ref{cplot}), opening the possibility of having a vanishing $C$ in the infinite volume
limit. In this sense, $C$ can be seen as a measure of the finite volume corrections to the pure power law.

In conclusion, the results reported in this section give support to an infrared gluon propagator
behaving as a pure power law, with
\begin{equation}
    D( q^2 ) \sim \left( q^2 \right)^{2 \kappa -1} \sim \left( q^2 \right)^{0.06} \, .
\end{equation}     
Accordingly, this means a vanishing zero momentum gluon propagator in agreement with the GZ
confinement mechanism.

%========================================================
%========================================================
%========================================================
%======================================================
%=====================================================
%=====================================================
\begin{table*}
\caption{\label{tablegluefit} 
Fitting the gluon ratios with equation (\ref{gluefit}) for $L^3 \times 256$
lattices. The first line is the maximum momentum used in the fit. 
$\chi^2$ stands for $\chi^2 / d.o.f.$. The errors in $\kappa$ are statistical 
and were computed with the bootstrap method.}
\begin{ruledtabular}
\begin{tabular}{c c @{\hspace*{-5mm}}dd  @{\hspace{-0mm}}dd  @{\hspace{-0mm}}dd  @{\hspace{-0mm}}dd  @{\hspace{-0mm}}dd }
  & $q_{max}:$ &  \multicolumn{2}{c}{\hspace*{7mm} 191 MeV \hspace*{0mm} } 
               &  \multicolumn{2}{c}{\hspace*{7mm} 238 MeV \hspace*{0mm} }
               &  \multicolumn{2}{c}{\hspace*{7mm} 286 MeV \hspace*{0mm} }
               &  \multicolumn{2}{c}{\hspace*{7mm} 333 MeV \hspace*{0mm} }
               &  \multicolumn{2}{c}{\hspace*{7mm} 381 MeV \hspace*{0mm} }   \\
\hline
  $L$ &  & \multicolumn{1}{c}{\hspace*{7mm}Param.}      &  \multicolumn{1}{r}{ $\chi^2$ }
         & \multicolumn{1}{c}{\hspace*{7mm}Param.}      &  \multicolumn{1}{r}{ $\chi^2$ }
         & \multicolumn{1}{c}{\hspace*{7mm}Param.}      &  \multicolumn{1}{r}{ $\chi^2$ }
         & \multicolumn{1}{c}{\hspace*{7mm}Param.}      &  \multicolumn{1}{r}{ $\chi^2$ }
         & \multicolumn{1}{c}{\hspace*{7mm}Param.}      &  \multicolumn{1}{r}{ $\chi^2$ }   \\
\hline
\hline
8   & $\kappa$ & 0.526(27)  & 0.12 & 0.531(19)  & 0.11 & 0.531(13)  & 0.08 & 0.522(16)  & 0.48 & 0.527(12)  & 0.54\\
    & $C$      & -0.179(54) &      & -0.194(34) &      & -0.193(19) &      & -0.171(28) &      & -0.184(18) & \\
\hline
10  & $\kappa$ & 0.511(35)  & 0.69 & 0.531(25)  & 0.98 & 0.525(21)  & 0.74 & 0.523(17)  & 0.56 & 0.527(16)  & 0.50 \\
    & $C$      & -0.114(66) &      & -0.161(42) &      & -0.146(30) &      & -0.144(21) &      & -0.150(19) &  \\
\hline
12  & $\kappa$ & 0.509(31)  & 0.11 & 0.517(21)  & 0.16 & 0.508(18)  & 0.33 & 0.521(18)  & 0.84 & 0.530(14)  & 1.03 \\
    & $C$      & -0.094(56) &      & -0.112(35) &      & -0.094(25) &      & -0.119(27) &      & -0.138(18) &  \\
\hline
14  & $\kappa$ & 0.536(24)  & 0.33 & 0.540(19)  & 0.20 & 0.548(16)  & 0.39 & 0.545(12)  & 0.34 & 0.542(11)  & 0.34\\
    & $C$      & -0.114(44) &      & -0.123(30) &      & -0.140(21) &      & -0.134(15) &      & -0.127(12) & \\
\hline
16  & $\kappa$ & 0.539(22)  & 1.77 & 0.528(17)  & 1.24 & 0.534(12)  & 0.96 & 0.536(12)  & 0.78 & 0.539(11)  & 0.68\\
    & $C$      & -0.125(43) &      & -0.102(30) &      & -0.112(19) &      & -0.118(14) &      & -0.123(12) & \\
\hline
18  & $\kappa$ & 0.529(20)  & 0.39 & 0.516(16)  & 0.77 & 0.523(14)  & 0.85 & 0.536(11)  & 1.79 & 0.5398(95) & 1.58 \\
    & $C$      & -0.099(36) &      & -0.068(25) &      & -0.085(19) &      & -0.111(14) &      & -0.119(13) & 
\end{tabular}
\end{ruledtabular}
\end{table*}
%=====================================================
%===================================================

%======================================================================
%======================================================================
\subsection{Finite Volume and Asymmetry Effects}

So far we have used only data from asymmetric lattice simulations with $T=256$. As a check to the
above results, in this subsection we investigate the finite volume effects and the changes introduced
from using a time-like direction much larger than the spatial directions. Therefore, besides the
asymmetric lattices of table \ref{Uvol}, we are going to consider the data coming from the lattices reported in table \ref{Ucheck}.

\begin{figure}[t]
\vspace*{0.2cm}                                  
\includegraphics[scale=0.35]{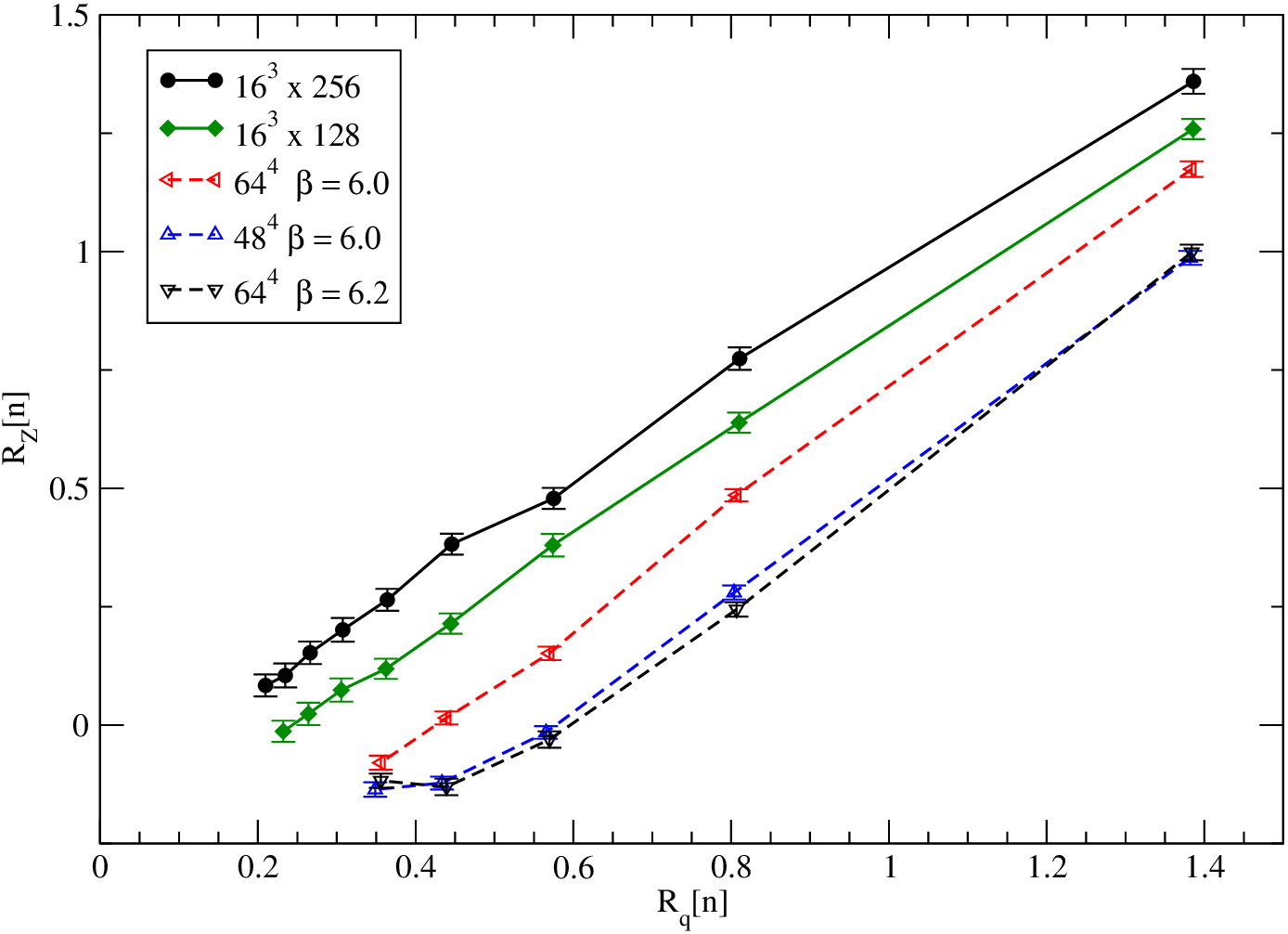}
\caption{\label{rcheck} $R_Z$ as function of $R_q$ for the lattices of table \ref{Ucheck} and
the two largest asymmetric lattices of table \ref{Uvol}. The figure contains data for $q< 500$ MeV for $16^3\times256$, $q< 900$ MeV for $16^3\times128$ 
, $q < 1.5$ GeV for $48^4$ and $\beta = 6.0$,
$q < 1.2$ GeV for $64^4$ and $\beta = 6.0$ and $q < 1.6$ GeV for $64^4$ and $\beta = 6.2$.}
\end{figure}

For on-axis momenta, the gluon propagator data (not shown here) for the lattices reported in tables
\ref{Uvol} and \ref{Ucheck} show very clearly finite volume and asymmetry effects. 
However, here we are interested in the ratios $R_Z$ as a function of $R_q$, which are reported in
figure \ref{rcheck}. 

Again, the data shows a linear behaviour for all the lattices of table \ref{Ucheck}. A fit of the ratios
computed using the $16^3 \times 128$ data to (\ref{gluefit}) for the smallest fitting range 
($q \leq 381$ MeV) gives $\kappa=0.541(19)$ and $C=-0.239(38)$ with a $\chi^2/dof=0.01$. 
The $\kappa$ value is in excellent agreement with the corresponding figure for $16^3\times 256$. 
For the $C$ figures, it comes that $C_{128}\simeq 2 \times C_{256}$. This result reinforces our interpretation of $C$ as a measure of the finite volume corrections to the pure power law.

In what concerns the symmetric lattice data in figure \ref{rcheck}, despite their relative large
smallest momentum, in the infrared the slope follows essentially the slope of the asymmetric ratios. 
The reader should be aware that, due to the $q_{min}$ of the symmetric lattices considered here, extracting infrared exponents by fitting the ratios for the symmetric lattices to equation (\ref{gluefit}) could be somewhat meaningless. For each of the symmetric lattices, one can estimate a $\kappa$
directly from the two most right points which gives $\kappa \sim 0.6$ or higher. The main
difference seems to be in the constant $C$, which takes larger values for smaller lattices.
In particular, note that $\beta = 6.2$, $64^4$ and $\beta = 6.0$, $48^4$ have essentially the same 
physical volume and their ratio data is, within statistical errors, indistinguishable. 

Given the results summarized in figure \ref{rcheck}, one can conclude that the ratios defined 
in (\ref{def_rat}) provide reliable estimates of the gluon infrared exponent, i.e. the
estimates seem to be independent of the lattice volume and asymmetry.

%========================================================
%========================================================
%========================================================
%======================================================

%=====================================================
%=====================================================
\begin{table}
\caption{Lattice setup used to check finite volume and asymmetry effects.
To convert to physical units we used the central values for the
lattice spacing reported in \cite{LatSpa}: $a = 0.1016$ fm for $\beta = 6.0$ and $a = 0.0728$ fm
for $\beta = 6.2$.  Volume is reported in fm. $q_{min}$ (in MeV) is the smallest non-vanishing momentum
for each lattice. For symmetric lattices only on-axis momenta are considered.}
\label{Ucheck}
\begin{ruledtabular}
\begin{tabular}{rlrrr}
  $\beta$  &   Lat  & Volume & $q_{min}$ & \# Conf. \\
\hline
    6.0  & $16^3 \times 128$  & $(1.63)^3 \times (13.00) $ & 95 & 164  \\
    6.0  & $48^4$                      & $(4.88)^4$ & 254 & 104  \\
    6.0  & $64^4$                      & $(6.50)^4$ & 190 &120  \\
\hline
    6.2  & $64^4$                      & $(4.66)^4$ & 266 & 99  \\
\end{tabular}
\end{ruledtabular}
\end{table}
%==================================================
%==============================================
%======================================================================
\section{The Ghost Propagator \label{ghost}}

%=======================================================
\begin{figure}[t]            
\vspace*{0.3cm}
\includegraphics[scale=0.33]{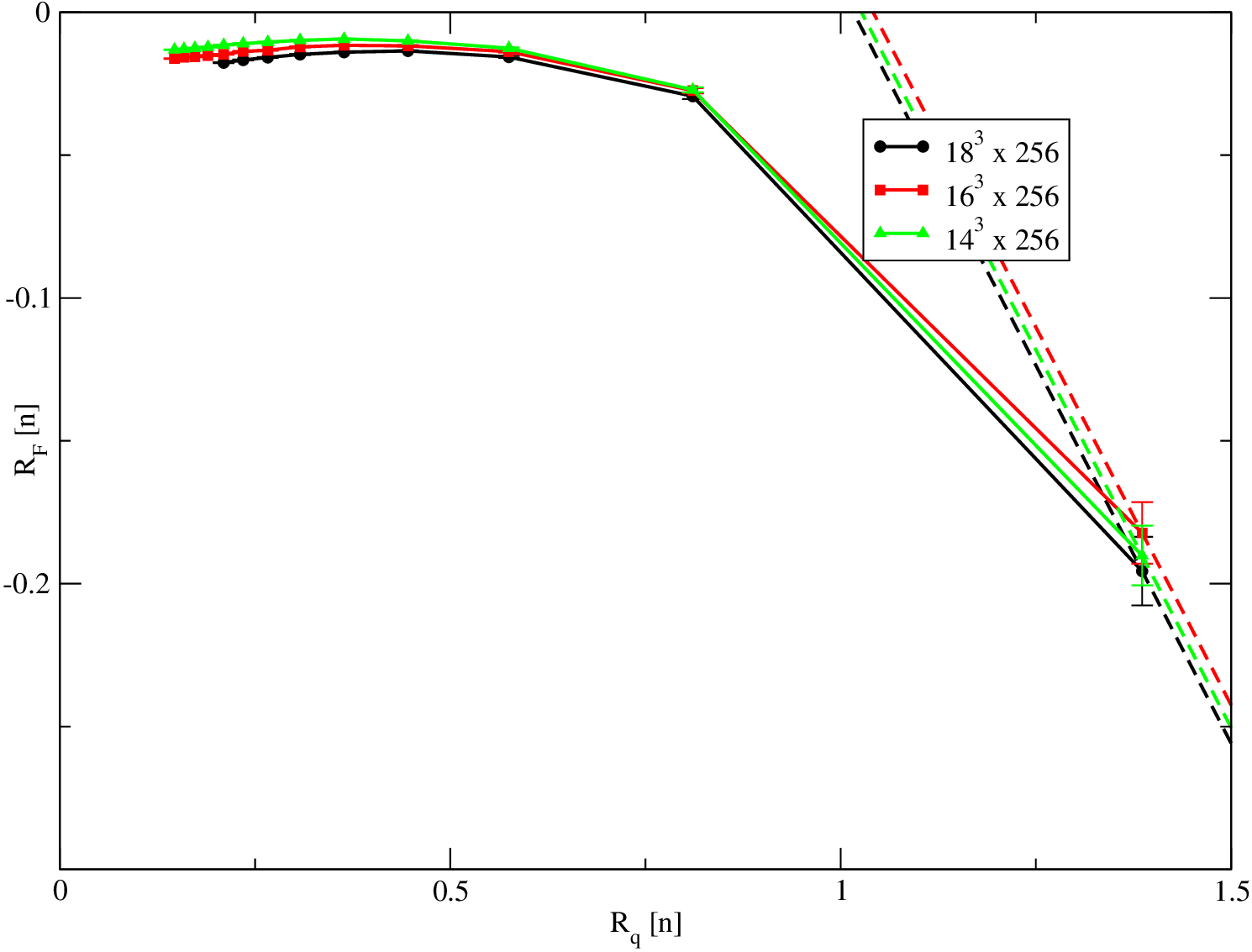}
\caption{\label{xplotghostratios} 
$R_{F}[n]$ as a function of $R_{q}[n]$
for the three largest 
lattices. The dash lines show, for each lattice, the curve 
$- \kappa \, R_{q}[n] \, + \, C$ where 
$\kappa = 0.529$ and $C$ adjusted to reproduce the right end point in the 
graph.}
\end{figure}
%===============================================

In what concerns the ghost propagator, the data for the ratios of the dressing functions, 
see figure \ref{plotratios}, do not show a linear behaviour as in the case of the gluon propagator. 
We have tried a number of functional forms to fit the data, but their $\chi^2/d.o.f.$ was always too large. 
At most, the slope of ratios of the ghost dressing function suggests a negative value for 
the ghost infrared exponent. 

In figure \ref{xplotghostratios} we show the ratios of the ghost dressing functions for the three larger 
lattices, including the curves
\begin{equation}
R_{F}[n] ~  = ~  - \kappa  R_{q}[n] ~ + ~ C ~,
  \label{GhostLinear}
\end{equation}
where $\kappa = 0.529$ and $C$ was adjusted to reproduce the ratio 
computed using our smallest momenta. The figure shows that either the data is still far from the linear behaviour or the infrared ghost propagator does not follow a pure power law.

%====================================================================
\section{Conclusions \label{conclusions}}

In conclusion, in this article we discuss a method which does not rely on extrapolations to the
infinite volume and i) allows to check if the infrared gluon and ghost propagators follow a pure power law; ii) measure the infrared exponents if the propagator is compatible with a power law.

For the gluon propagator the lattice data validates the power law behavior and the corresponding
infrared exponent is measured. From the two large asymmetry lattices and the two smallest fitting
ranges we have $\kappa = 0.529(20)$ and $0.539(22)$ which gives a combined value of
$\kappa = 0.535(14)$. As discussed, the ratio method devised in this paper seems to be independent
of the asymmetry and lattice volume, at least for the simulations reported here. This gives confidence
on our estimate for $\kappa$. Furthermore, the reported exponent $\kappa = 0.535(14)$ is within the
range of values allowed by the renormalization group analysis of  \cite{Pa04,FiGi04}. Given the results
for the ghost propagator, it is not clear to the authors if this $\kappa$ should be or not in that
range. More, the 
%PAULO
measured 
%measure
$\kappa$ supports a vanishing zero momentum gluon propagator in
agreement with the Gribov-Zwanziger gluon confinement scenario.

The reader should note that, on a finite lattice, simulations have produced always a finite and 
non-zero value for the gluon propagator at zero momentum. The method discussed here do not
use directly $D(0)$. Anyway, as mentioned before, the conclusions are in perfect agreement with
the scaling analysis of $D(0)$ as a function of the volume performed in \cite{OlSi08}, which gives
a $D(0)=0$ in the infinite volume limit. In this sense, the results of \cite{OlSi08} reconciliate the
results of the lattice simulations with those reported in this work, i.e. that in the Landau gauge
the continuum result is $D(0) = 0$.

In what concerns the ghost propagator, the lattice data suggests that the propagator does not follow a
pure power law. In this sense, our simulation does not validate the infrared solution  (\ref{Fdse}) of the
DSE. The observed difference between the lattice and the DSE solution could be because either the
lattices used here are not long enough or the ghost propagator does not follow a pure power law in 
the deep infrared region. The answer to this question requires further investigations.

%====================================================================
\begin{acknowledgments}
This work was supported in part by F.C.T. under contracts
POCI/FP/63436/2005 and POCI/FP/63923/2005. 
P.J.S. acknowledges financial support from FCT via 
grant SFRH/BD/10740/2002.
\end{acknowledgments}

\appendix
\section{Finite Volume Effects on the Infrared Gluon Dressing Function: an effective parameterization}

The results on the ratios of the gluon dressing function, discussed in section \ref{ratiosglsec}, suggest
that the finite volume and asymmetry corrections are summarized in $C$.  
Let $\Delta (q)$ be the multiplicative correction to the continuum dressing function $Z(q^2)$, i.e let us
assume that the lattice dressing function is given by
\begin{equation}
   Z_{Lat} (q^2) ~ = ~ Z(q^2) \, \Delta(q) \, .
\end{equation}
Then, $\Delta (q[n+1]) =  \Delta (q[n]) e^C$ which allows to write
\begin{eqnarray}
  \frac{d \Delta (q)}{d q} & \sim & 
         \frac{\Delta (q[n+1]) - \Delta(q[n])}{q[n+1] - q[n]} 
    \nonumber \\
  & \sim & \Delta (q) \, \frac{e^C - 1 }{\frac{2 \pi}{a T}} 
    ~ = ~ \Delta (q) \, A
\end{eqnarray}
where $A$ is a constant. The integration of the last equation gives
$\Delta (q) ~ = ~ \Delta_0 \, \exp (A q)$ 
where $\Delta_0$ is a constant of integration. 
The reader should note that this form for $\Delta (q)$ is only valid in the infrared region.
Computing $\Delta (p)$ for the full range of momenta is beyond the scope of this work.
Nevertheless, for high momenta, where the propagator is well described by its perturbative
behavior and the lattice gluon propagator does not seem to depend on the volume, one should 
have $\Delta(p) \sim 1$. In principle, this restriction allow us to fix $\Delta_0$ via the normalization
condition $\Delta(q_h)=1$ where $q_h$ is a sufficiently high momentum value. The present lack of
knowledge on the functional form for $\Delta(p)$ at momenta $q > 400$ MeV prevent us to compute
$\Delta_0$. At best, one can estimate $\Delta_0$ at the highest momentum $q$ ($\sim 400$ MeV)
where one can still approximate $\Delta (q) ~ = ~ \Delta_0 \, \exp (A q)$. Then,
 $\Delta_0=\exp(-A q_h)$ and 
\begin{equation}
\Delta (q) ~ = ~ \Delta_0 \, \exp (A q) ~ \sim ~  \, \exp \big( A (q-q_h) \big)
\label{delta}
\end{equation}
for the infrared region.

For practical purposes, like fitting the lattice data,  $\Delta_0$ can be absorbed into the definition of 
$\omega$, and we get an exponential correction
to $Z(q^2)$,
\begin{equation}
  Z_{Lat} (q^2) ~ = ~ \omega \left( q^2 \right)^{2 \kappa} \, e^{A q}  \, ,
  \label{ZCorrLat}
\end{equation}
with the constant $A$ parametrizing the finite volume effects and/or the corrections to the pure power
law and $\kappa$ is the continuum exponent. The results of fitting (\ref{ZCorrLat}) to 
the lattice gluon dressing function are reported in table \ref{tablegluefitCorr}.

The $\kappa$ values in tables \ref{tablegluefit} and \ref{tablegluefitCorr} are essentially the same. 
The good agreement between the two results can be viewed as a cross-check of the original method
discussed previously and, in this sense, gives further confidence in the final result. The reader should note that $A$, like $C$, takes negative values and approaches zero as the lattice volume increases.

\begin{figure}[h]
\vspace*{0.2cm}                                  
\includegraphics[scale=0.35]{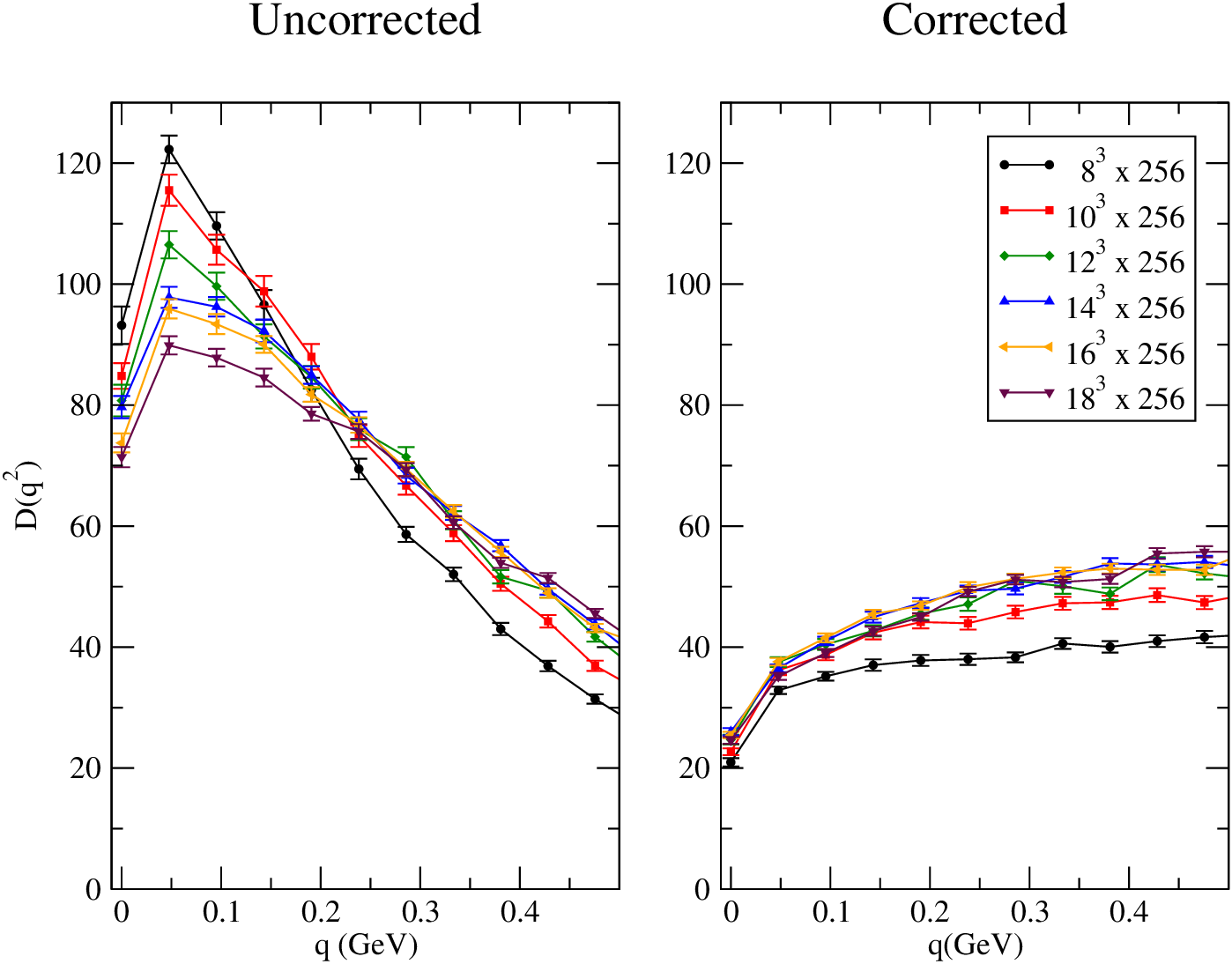}
\caption{\label{corrglue} Corrected and uncorrected gluon propagator using eq. (\ref{delta}) with $q_h=400$ MeV. }
\end{figure}

In what concerns the function $\Delta(q)$, see eq. (\ref{delta}), it provides a correction to the gluon
propagator which decreases with increasing momenta, as expected if it models the finite volume 
effects. Indeed, $\Delta(q=0) \sim \exp (-A q_h) \gg 1$ and  $\Delta(q_h) \sim 1$  -- see figure \ref{corrglue} where we have corrected the propagator using $q_{h}=400 MeV$. 
The reader should not forget that we are only estimating the corrections due to finite volume/asymmetry.
Figure \ref{corrglue} shows a corrected gluon propagator which is supressed in the infrared region.
Moreover, the corrected data shows also a good agreement between all propagators, excluding our
smallest lattice, for momenta below 200 MeV.

%========================================================
%========================================================
%========================================================
%===================================================

%=====================================================================
\begin{table*}
\caption{\label{tablegluefitCorr} 
Fitting the gluon dressing functions to (\ref{ZCorrLat}) for $L^3 \times 256$
lattices. The first line is the maximum momentum used in the fit. $\chi^2$ 
stands for $\chi^2/d.o.f.$. The errors are statistical and were computed with 
the bootstrap method. A in GeV$^{-1}$.} 
\begin{ruledtabular}
\begin{tabular}{c c @{\hspace*{-5mm}}dd  @{\hspace{-0mm}}dd  @{\hspace{-0mm}}dd  @{\hspace{-0mm}}dd  @{\hspace{-0mm}}dd }
\hline
  & $q_{max}:$  &  \multicolumn{2}{c}{\hspace*{7mm} 191 MeV \hspace*{0mm} } 
               &  \multicolumn{2}{c}{\hspace*{7mm} 238 MeV \hspace*{0mm} }
               &  \multicolumn{2}{c}{\hspace*{7mm} 286 MeV \hspace*{0mm} }
               &  \multicolumn{2}{c}{\hspace*{7mm} 333 MeV \hspace*{0mm} }
               &  \multicolumn{2}{c}{\hspace*{7mm} 381 MeV \hspace*{0mm} }   \\
\hline
  $L$ &  & \multicolumn{1}{c}{\hspace*{7mm}Param.}      &  \multicolumn{1}{r}{ $\chi^2$ }
         & \multicolumn{1}{c}{\hspace*{7mm}Param.}      &  \multicolumn{1}{r}{ $\chi^2$ }
         & \multicolumn{1}{c}{\hspace*{7mm}Param.}      &  \multicolumn{1}{r}{ $\chi^2$ }
         & \multicolumn{1}{c}{\hspace*{7mm}Param.}      &  \multicolumn{1}{r}{ $\chi^2$ }
         & \multicolumn{1}{c}{\hspace*{7mm}Param.}      &  \multicolumn{1}{r}{ $\chi^2$ }   \\
\hline
\hline
8  & $\kappa$ & 0.526(26)   & 0.09 & 0.533(19) & 0.12 & 0.534(11) & 0.08 & 0.523(10) & 0.62 & 0.524(9)  & 0.51\\
   & $A$      & -3.75\pm1.1 &      & -4.06(68) &      & -4.11(34) &      & -3.69(28) &      & -3.73(23) & \\
\hline
10 & $\kappa$ & 0.511(27)   & 0.53 & 0.536(22) & 1.08 & 0.534(17) & 0.73 & 0.531(14) & 0.58 & 0.534(13) & 0.49 \\
   & $A$      & -2.3\pm1.1  &      & -3.40(69) &      & -3.33(51) &      & -3.22(37) &      & -3.30(29) &  \\
\hline
12 & $\kappa$ & 0.508(31)   & 0.07 & 0.515(22) & 0.12 & 0.507(15) & 0.24 & 0.520(12) & 0.84 & 0.537(9)  & 1.94 \\
   & $A$      & -1.9\pm1.2  &      & -2.25(78) &      & -1.92(46) &      & -2.40(36) &      & -2.96(23) &  \\
\hline
14 & $\kappa$ & 0.538(23)   & 0.24 & 0.542(18) & 0.17 & 0.552(14) & 0.47 & 0.551(11) & 0.36 & 0.546(9)  & 0.45  \\
   & $A$      & -2.42(87)   &      & -2.62(59) &      & -3.00(41) &      & -2.96(29) &      & -2.80(21) & \\
\hline
16 & $\kappa$ & 0.541(22)   & 1.15 & 0.532(16) & 0.78 & 0.535(10) & 0.55 & 0.539(9)  & 0.50 & 0.543(8)  & 0.54\\
   & $A$      & -2.67(84)   &      & -2.29(54) &      & -2.39(31) &      & -2.53(24) &      & -2.66(18) & \\
\hline
18 & $\kappa$ & 0.529(20)   & 0.28 & 0.516(15) & 0.59 & 0.523(12) & 0.54 & 0.539(9)  & 2.14 & 0.550(8)  & 2.71 \\
   & $A$      & -2.05(79)   &      & -1.50(51) &      & -1.75(33) &      & -2.31(24) &      & -2.66(20) & \\
\hline
\end{tabular}
\end{ruledtabular}
\end{table*}
%===============================================%
%=========================================================
%\vspace*{-0.6cm}


\begin{thebibliography}{99}
\bibitem{su2vsu3} 
 Note that in \cite{su2vsu307} and \cite{su2vsu3adelaide} the pure gauge propagators are compared for the SU(2) and SU(3) gauge theories. The results from the two groups suggest that $D(q^2)$ and $G(q^2)$ are independent of the gauge group for the all range of momenta.
\end{thebibliography}
\end{document}